\documentclass[10pt,aps,rmp,onecolumn,superscriptaddress,floatfix,nofootinbib,showkeys,notitlepage,preprintnumbers]{revtex4-1}
\usepackage{amsmath,amssymb,amsfonts}
\usepackage{graphicx,color}
\usepackage{natbib}
\usepackage{eurosym}
\usepackage{hyperref}
\usepackage[inline]{enumitem}
\usepackage{soul}
\newcommand{\ie}{\emph{i.e.}}
\newcommand{\eg}{\emph{e.g.}}

\begin{document}

\title{How the interbank market becomes systemically dangerous: an agent-based network model of financial distress propagation}

\author{Matteo Serri}\email{matteo.serri@gmail.com}
\affiliation{\small Universit\`a ``Sapienza'' Facolt\`a di Economia (Dip. MEMOTEF) - 00161 Rome (Italy)}
\author{Guido Caldarelli}
\affiliation{\small IMT School for Advanced Studies - 55100 Lucca (Italy)}
\affiliation{\small Istituto dei Sistemi Complessi (ISC)-CNR - 00185 Rome (Italy)}
\affiliation{\small London Institute for Mathematical Sciences - Mayfair W1 London (UK)}
\author{Giulio Cimini}
\affiliation{\small IMT School for Advanced Studies - 55100 Lucca (Italy)}
\affiliation{\small Istituto dei Sistemi Complessi (ISC)-CNR - 00185 Rome (Italy)}

\begin{abstract}
Assessing the stability of economic systems is a fundamental research focus in economics, that has become increasingly interdisciplinary in the currently troubled economic situation. 
In particular, much attention has been devoted to the interbank lending market as an important diffusion channel for financial distress during the recent crisis. 
In this work we study the stability of the interbank market to exogenous shocks using an agent-based network framework. 
Our model encompasses several ingredients that have been recognized in the literature as pro-cyclical triggers of financial distress in the banking system: 
credit and liquidity shocks through bilateral exposures, liquidity hoarding due to counterparty creditworthiness deterioration, 
target leveraging policies and fire-sales spillovers. But we exclude the possibility of central authorities intervention. 
We implement this framework on a dataset of 183 European banks that were publicly traded between 2004 and 2013. 
We document the extreme fragility of the interbank lending market up to 2008, when a systemic crisis leads to total depletion of market equity with an increasing speed of market collapse. 
After the crisis instead the system is more resilient to systemic events in terms of residual market equity. 
However, the speed at which the crisis breaks out reaches a new maximum in 2011, and never goes back to values observed before 2007. 
Our analysis points to the key role of the crisis outbreak speed, which sets the maximum delay for central authorities intervention to be effective.
\end{abstract}
\keywords{Financial contagion; Systemic risk; Interbank lending market; Agent-based models}

\maketitle

\section{Introduction}

The financial instability which characterised the last decade made clear to academic and regulators that the economy and the financial system have become so inherently complex that a multidisciplinary effort 
is needed to disentangle the intertwined set of connections between different actors and institutions \citep{Gai2011,Acemoglu2013,Sorkin2009,Beale2011,Haubrich2013}. 
Indeed, the network structure of the financial system is now acknowledged as a potential trigger of instability \citep{Lau2009,Bardoscia2016}, thus the many recent studies on the origin of the crisis 
investigating the interplay between network topology and contagion processes (an approach originally developed in statistical physics\footnote{Net gains. {\em Nature Physics} {\bf 9}, 119 (2013) doi:10.1038/nphys2588.}). 
Among the various subjects, researchers focused in particular on the interbank lending market, namely the network of financial interlinkages between banks resulting from unsecured overnight loans. 
This system has been ascribed as one of the principal diffusion channels for financial distress during the 2007/2008 crisis \citep{Bluhm2011,Cont2013,Gabrieli2014,Georg2013,Krause2012}: 
after the collapse of Lehman Brothers, the interbank market froze causing a severe liquidity drought within the whole financial system \citep{Acharya2010,Adrian2009,Brunnermeier2009a,Berrospide2012}. 
Such a black swan was due to the collapse of different liquidity channels, \eg, Asset-Backed-Commercial-Paper and Repo, and to a burst in the spread between long-term and over-night interest rates (or Libor-OIS spread) 
\citep{Brunnermeier2009a}. As this problem became systemic, central banking authorities intervened with extraordinary monetary policies to restore the solvency of the financial system. 

Liquidity issues threaten the stability of the financial system by generating important spillover effects. 
Different subcategories of problems have been identified in the literature: funding liquidity, market liquidity, flight to liquidity-quality, liquidity spirals, liquidity hoarding, market freeze, and assets fire sales. 
Authors like \citet{Skeie2011} and \citet{Brunnermeier2009b} have modeled liquidity dynamics using a theoretical approach. Others like \citet{Berrospide2012} and \citet{Acharya2010} 
resorted to empirical econometric analyses to study the causes of interest rate spreads. \citet{Eisenberg2001} have been the first to tame the complexity of the problem using a theoretical approach 
explicitly considering the set of interconnections within the financial system. Their work originated a flourishing line of research aimed at assessing the systemic importance of financial institutions 
under a network perspective---see, among the others, \citet{Greenwood2015,Gabrieli2014,Bluhm2011,Hausenblas2015,Barucca2016}.

A second approach to deal with the complexity of the financial system has been that of using agent-based models (ABM). 
An ABM is a simulated framework in which several agents interact following optimal selfish strategies, creating spin-off effects such as the emergence of an endogenous trading market \citep{Caporale2007,Lucas2014}. 
The use of ABM in economics and finance started in parallel with the development of calculators and computer science. 
The Santa Fe Institute Artificial Stock Market model in the early 1990s was one of the first ABM developed---and later complemented with market orders \citep{Lux1998}. 
Recent advances in ABM for financial stability studies include the work of \citet{Fischer2013}, who showed the fundamental role of leverage in assessing systemic risk, and of \citet{Georg2013} and \citet{Halaj2014}, 
who modeled an emerging interbank market via a stylised trading mechanisms. All these studies agree on the relevance of the topology of interactions for contagion mechanisms. 
Others studies like \citet{Cont2013}, \citet{Bookstaber2014} and \citet{Klinger2014} consider an exogenous interbank network (data-driven or simulated) affected by shocks that induce an idiosyncratic response 
such as the emergence of bank runs and fire sales. The aim is to evaluate systemic risk, and find effective regulatory capital buffer and requirements to prevent cascade failures.

In this work we bring together these two approaches by introducing and ABM which describes the network dynamics of the interbank market. 
We build on the framework introduced by \citet{Lau2009} and \citet{Krause2012}, and later developed by \citet{CiminiSerri2016}. 
We consider a data-driven network of bilateral exposures between banks (the agents of our model), which use micro-optimal rules to interact with each other and with the rest of the financial system. 
We explicitly model various categories of spillover effects arising during financial crises, such as fire sales and interest rate jumps due to leverage targeting and liquidity hoarding behavior of banks. 
The modeled dynamics of pro-cyclical policies then spread financial losses via credit and liquidity interconnections, and may result in cascades of defaults. 
In our approach, we just assume the existence of events in order to focus on the description of the dynamical evolution of the financial system. 
Our ABM can thus be used to stress-test the robustness of the financial system to an external shock, which can be either absorbed or cause an avalanche of failures eventually leading to the market freezing.

Note that the use of an ABM allows us to have a complete description of the system dynamics during a crisis (\ie, out of the economic equilibrium), which would be very difficult to obtain by analytical modeling. 
The ABM presented here also allows to consider a flow of events which is different from what happened in reality during, \eg, the 2007-2008 financial crisis. 
In particular, we are interested in the scenario characterised by the absence of a lender of last resort like a central bank, whose monetary policies can completely re-design the market. 
Indeed, our aim is not to reproduce the real dynamics of the crisis, but to define the worst-possible scenario without any quantitative easing nor bail-out program by regulatory institutions. 
The rest of the paper is organised as follows. Sections II and III report basic assumptions and detailed description of the ABM framework, respectively. 
Results of our extensive simulation program are discussed in Section IV, and Section V concludes.

\section{Model assumptions}

The main ingredients of the model are the set of connections and the strategies of the agents. In order to define them, we make the choices listed below.

\subsection*{Network definition}
\begin{itemize}
\item The interbank network is assumed to consist of loans with overnight (ON)  duration. Thus, $A_{ij}$ is the overall amount that bank $i$ lends to bank $j$ (\ie, the interbank asset of $i$ towards $j$), 
which corresponds to the liquidity $L_{ji}\equiv A_{ij}$ that $j$ borrowed from $i$ (\ie, the interbank liability of $j$ towards $i$). As contracts are of short duration (ON), 
we assume that links are continuously placed and immediately resolved and rolled over, so that the same (current) interest rates $r>1$ applies to both assets and liabilities. 
In other words, the market dynamics we consider here is on a time scale longer than that of contracts duration.
\item The network is derived from aggregate interbank exposures and obligations: $A_i=\sum_j A_{ij}$ and $L_i=\sum_j L_{ij}\equiv\sum_j A_{ji}$. 
We use the Bureau van Dijk Bankscope database\footnote{Raw Bankscope data are available from Bureau van Dijk: https://bankscope.bvdinfo.com. Refer to~\cite{Battiston2015X} for all the details 
about the handling of missing data.}, that contains yearly-aggregated balance sheets information of $N=183$ large European banks from 2004 to 2013, 
and resort to the procedure described in \citet{Cimini2015} which uses the fitness model \citep{Caldarelli2002} to build an ensemble of interbank networks from such aggregate data. 
\item For each bank $i$, the balance sheet equation reads:
\begin{equation}\label{eq.BS}
E_i:=A_i^E-L_i^E+r\sum_jA_{ij}-r\sum_jL_{ij},
\end{equation}
where $A_i^E$ and $L_i^E$ are, respectively, the external assets and liabilities of $i$. 
$A_i=A_i^E+r\sum_jA_{ij}$ and $L_i=L_i^E+r\sum_jL_{ij}$ are the total amount of assets and liabilities (external plus interbank) held by $i$, respectively. 
For each bank $i$ to be solvent, it must be $E_i>0$. 
\end{itemize}

\subsection*{Strategies definition}
In order to build agent strategies and model dynamics, we take inspiration from the most important facts characterising the crises.
\begin{itemize}
\item If hit by a shock, a bank sells assets following a leverage targeting policy in order to reinforce its reputation and expectation of the stakeholders. 
\item After the shock and during the realignment, worries about creditworthiness may cause a ``flight to quality'',  for which banks withdraw liquidity from the market. 
\item Liquidity hoarding coupled with a constant liquidity demand triggers an increase of interbank interest rates, and the consequent revaluation of interbank assets and liabilities. 
\item If a bank defaults, credit and funding shocks propagate through its bilateral exposures like a bank-run contagion on financial interbank contracts.  
\item Interbank network connections and fire sales spillovers may lead to default cascades, with a consequent increasing of liquidity hoarding and interest rate. 
\item In extreme conditions, the market freezes triggering exacerbated fire sales.
\end{itemize}

\section{Model dynamics}

Building on the above definitions and assumptions, we now specify the model dynamics of the interbank market.

\subsubsection{Exogenous shock}

\begin{itemize}
\item At a given time step $t=t_0$, bank $s$ is hit by an exogenous shock, so that its external assets $A_s^E$ decrease by a quantity $\Phi$ \citep{Lau2009,Krause2012}:
\begin{equation}\label{eq.init_s}
A_s^E(t_0+1)=A_s^E(t_0)-\Phi\Rightarrow E_s(t_0+1)=E_s(t_0)-\Phi 
\end{equation}
\item At first, bank $s$ tries to realign to its target leverage $B_s(t_0)=A_s(t_0)/E_s(t_0)$ by selling assets. To this end, the amount of assets to be sold is given by \citep{Brunnermeier2009a,Adrian2009}:
\begin{equation}\label{eq.targ_lev}
A_{s}(t_0+1)-E_s(t_0+1)B_s(t_0)=A_{s}(t_0)-\Phi-[E_s(t_0)-\Phi]\,B_s(t_0)=\Phi[B_s(t_0)-1].
\end{equation}
As bank $s$ has increased worries on its financial situation, it adopts a microprudential policy \citep{Acharya2010,Berrospide2012} by hoarding the liquidity granted by such sales. 
This means that its interbank loans are not rolled over for their entire amount. We thus assume that external and interbank assets are sold proportionally to their balance sheet shares 
$f^E_s(t_0)=A_s^E(t_0)/A_s(t_0)$ and $f^I_s(t_0)=\sum_kA_{sk}(t_0)/A_s(t_0)$, respectively. We further assume that interbank assets rescale proportionally to the contracts size: 
each loan $A_{sk}(t_0)$ decreases of an amount equal to $\Phi[B_s(t_0)-1]f^I_s(t_0)$ times the ratio of $A_{sk}(t_0)$ itself to the total exposure $\sum_kA_{sk}(t_0)$ of $s$. 
The net result is that part of the total value of the interbank market is lost. The consequent increase of counterparty and roll-over risk perceived in the market 
causes the interbank interest rate to grow up \citep{Skeie2011}. In particular we assume:
\begin{equation}\label{eq.smallIR}
dr/dt=r\,\ln(1+\alpha)\Longrightarrow r(t_0+1)=(1+\alpha)\,r(t_0)+\varepsilon
\end{equation}
where the factor $\alpha>0$ is as a small quantity which leaves the system stable, and $\varepsilon$ is a random variable drawn from $\mathcal{N}[0,\sigma]$. 
Thus, interbank assets and liabilities increase as $A_{jk}(t_0+1)=[r(t_0+1)/r(t_0)]A_{jk}(t_0)$ and $L_{jk}(t_0+1)=[r(t_0+1)/r(t_0)]L_{jk}(t_0)$, respectively and $\forall j,k$. 
Note that while the external assets sold by $s$ are turned into liquidity\footnote{Models usually assume that external assets do not revalue as the most unbiased assumption that the overall contribution 
of market fluctuations averages to zero.} and thus do not cause any change of equity for $s$, the liquidated interbank assets causes the bank's equity to shrink as they do not get revalued by the new interest rate.  
Therefore, the target leverage of $s$ is substantially respected, except for the non appreciation of a part of its interbank assets (that is, however, minimal). Overall, the balance sheet of bank $s$ becomes:
\begin{eqnarray}\label{eq.balance_s}
E_s(t_0+1)&=&\Big\{A_s^E(t_0)-\Phi-\Phi[B_s(t_0)-1]f^E_s(t_0)\Big\}-\Big\{ L_s^E(t_0)-\Phi[B_s(t_0)-1]f^E_s(t_0)-\Phi[B_s(t_0)-1]f^I_s(t_0)\Big\} \nonumber \\
&+&[r(t_0+1)/r(t_0)]\sum_kA_{sk}(t_0)\left\{1-\frac{\Phi[B_s(t_0)-1]f^I_s(t_0)}{\sum_kA_{sk}(t_0)}\right\}-[r(t_0+1)/r(t_0)]\sum_kL_{sk}(t_0)= \\
&=&E(t_0)-\Phi+[\alpha+\varepsilon/r(t_0)]\left\{\sum_k[A_{sk}(t_0)-L_{sk}(t_0)]-\Phi[B_s(t_0)-1]f^I_s(t_0)\right\}, \nonumber 
\end{eqnarray}
indicating that the equity of $s$ has changed due to the external shock and the revaluation of its interbank contracts, except the part which is not rolled-over. 
\item Banks $\{i\}$ that borrow from $s$ receive a funding shock given by the interbank assets $s$ dries up for liquidity hoarding, and replace it with an external liability. Thus their balance sheets becomes:
\begin{equation}\label{eq.balance_k}
E_i(t_0+1)=E_i(t_0)+[\alpha+\varepsilon/r(t_0)]\left\{\sum_k[A_{ik}(t_0)-L_{ik}(t_0)]+\Phi[B_s(t_0)-1]\frac{L_{is}(t_0)}{A_s(t_0)}\right\}.
\end{equation}
\item For all other banks $\{j\}$:
\begin{equation}\label{eq.balance_j}
E_j(t_0+1)=E_j(t_0)+[\alpha+\varepsilon/r(t_0)]\left\{\sum_k[A_{jk}(t_0)-L_{jk}(t_0)]\right\}.
\end{equation}
\end{itemize}
These steps are repeated until a first default is triggered.

\subsubsection{Cascade Failures}

\begin{itemize}
\item After some rounds of exogenous shocks, at iteration step $t^*$ a given bank $u$ becomes insolvent and defaults, meaning $E_u(t^*)\le0$. Bank $u$ is removed from the system, 
but this event triggers a cascade of credit and liquidity losses in the interbank market \citep{Lau2009,Krause2012}. We use a one-step Debt-Solvency rank dynamics \citep{CiminiSerri2016} to model this process, \ie, we have 
\begin{enumerate}[label=\Alph*)]
 \item {\em Credit shocks}. Bank $u$ cannot meet its obligations, so that each other bank $j$ suffers a loss equal to $\lambda A_{ju}(t^*)$. 
 Here $\lambda$ indicates the amount of loss given default. We set $\lambda=1$ to consider only un-collateralized markets. 
 \item {\em Funding shocks}. Banks are unable to replace all the liquidity previously granted by the defaulted institutions but by selling their assets, triggering fire sales \cite{Brunnermeier2009b}. 
 In particular, each bank $j$ is able to replace only a fraction $(1-\rho)$ of the lost funding from $u$, and its assets trade at a discount: 
 $j$ must sell assets worth $[1+\gamma(t^*)]\rho A_{uj}(t^*)$ in book value terms\footnote{Following a common approach \citep{Ellul2011,Feldhutter2012,Greenwood2015}, we assume that fire sales 
 generate a linear impact on prices. Given that $Q(t^*)=\rho\sum_{j\neq u}A_{uj}(t^*)$ is the total amount of assets to be liquidated, we have no price change when $Q(t^*)=0$, 
 and assume that assets value vanishes when $Q(t^*)=C\equiv\sum_{ij}A_{ij}$ (\ie, when the whole market has to be sold). Thus we have a relative assets price change $\Delta p/p|_{t^*}=-Q(t^*)/C$. 
 To obtain the corresponding $\gamma(t^*)$, we equate the loss $\gamma(t^*)\rho A_{uj}(t^*)$ to the amount sold $[1+\gamma(t^*)\rho A_{uj}(t^*)$ times $\Delta p/p|_{t^*}$, 
 obtaining $\gamma(t^*)=[C/Q(t^*)-1]^{-1}$.}, corresponding to an overall loss of $\gamma(t^*)\rho A_{uj}(t^*)$. Here we set $\rho=1$, meaning that banks actually cannot replace the lost funding from $u$ 
 and are thus forced to entirely replace the corresponding liquidity by assets sales. 
\end{enumerate}
Overall, the balance sheet of a bank $j$ connected to $u$ becomes:
\begin{equation}\label{eq.balance_after_def}
E_j(t^*+1)=E_j(t^*)-\lambda A_{ju}(t^*)-\gamma(t^*)\rho A_{uj}(t^*).
\end{equation}
If any other bank $u'$ fails because of the suffered loss, the procedure above is repeated, until no other bank fails.
\item After the default cascade has ended, the net change of equity drives each survived bank to realign to its leverage before the cascade, and to liquidate some of its assets. 
Thus, the dynamics restart from the exogenous shock phase, even if the shock is endogenous this time. However, now the interbank market has shrunk significantly by a loss 
$\Delta E(t^*)=\sum_iE_i(t^*+1)-E_i(t^*)\le0$. This triggers a sudden interest rate increase, which we model by adding to eq. (\ref{eq.smallIR}) 
a source term depending on the ratio of $\Delta E(t^*)$ to the exogenous shock $\Phi$:
\begin{equation}\label{eq.bigIR}
dr/dt=r\,\ln(1+\alpha)+\delta\ln[|\Delta E(t^*)|/\Phi]\Longrightarrow r(t^*+1)=(1+\alpha)\,r(t^*)+\alpha\,\delta\,\log_{[\alpha+1]}[|\Delta E(t^*)|/\Phi]+\varepsilon.
\end{equation}
Thus, if $|\Delta E(t^*)|\simeq\Phi$ the interest rate grows by the same factor $\alpha$ as before, but if $|\Delta E(t^*)|\gg\Phi$ the interest rate blows up. 
Overall, the equities of each bank $s$ change\footnote{We consider a random sequence of survived banks to perform target leveraging sequentially.} according to eq. (\ref{eq.balance_s}) with $t_0=t^*$ , 
where however $\Phi$ is replaced by $E_s(t^*)-E_s(t^*+1)$, and the interest rate has changed according to eq. (\ref{eq.bigIR}). 
This process may trigger again a default cascade. Otherwise, the exogenous shock dynamics continues afterwards.
\end{itemize}

\subsubsection{Market freeze}

The market freezes at iteration $t=t_c$ when the total relative equity of the market $\sum_iE_i(t_c)/\sum_iE_i(0)$ becomes smaller than a critical ratio $\epsilon_c$. 
At this point, interbank assets get totally liquidated. Whenever for a given bank $j$ this liquidation is not enough to repay debts (that is, if $\chi_j(t_c)=\sum_k[L_{jk}(t_c)-A_{jk}(t_c)]>0$), 
such bank is forced to fire sale its external assets. However, differently from normal sales due to funding shocks, now the market is frozen and the value of external assets is therefore enormously decreased.  
Thus bank $j$ must sell a fraction of assets worth $(1+\Gamma_c)\chi_j(t_c)$, with $\Gamma_c\gg\gamma(t_c)$. 
To evaluate the depricing factor $\Gamma_c$, we rescale $\gamma(t_c)$ by the relative wealth of potential buyers of fire-sold assets \cite{Duarte2015} 
(\ie, their current wealth compared to the initial value):
\begin{equation}\label{eq.fire_sale}
\Gamma_c=\gamma(t_c)\,\frac{\sum_iE_i(0)}{\sum_i\{E_i(t_c)-\chi_i(t_c)\Theta[\chi_i(t_c)]\}},
\end{equation}
meaning that if the interbank market shrinks it is more difficult to sell assets and $\Gamma_c$ grows. Note that to compute $\Gamma_c$, 
we subtract from the current total equity the total assets that must be fire sold (because these assets cannot be used to acquire other assets). 
The market freeze condition ends the ABM dynamics of the interbank market.

\section{Results}

\begin{table}[h!]
\caption{List of parameter values used in simulations.}
\begin{tabular}{c|c|l} 
\hline						
{\bf Symbol}	&	{\bf Value}	&	{\bf Description} 	\\
\hline\hline
$d$	& $0.1$	&	density of the reconstructed interbank network \\
$\lambda$ & 1.0 &	loss given default \\
$\rho$ & $1.0$	& 	lost funding fraction to be replaced \\
$\Phi$ & $10^8$ \$ &	entity of the exogenous shock \\
$r_0$ & $1.0$ &		initial interest rate \\
$\alpha$ & $10^{-3}$ &	interest rate increase factor \\
$\sigma$ & $10^{-3}$ &	variation of the random variable in the interest rate dynamics \\
$\delta$ & $10^{-2}$ &	prefactor of the source term for interest rate dynamics \\
$\epsilon_c$ & $0.37$	& critical ratio of residual equity for market freeze \\
\hline
\end{tabular}
\label{tab.params}
\end{table}

\begin{figure}[h!]
\centering
\includegraphics[width=\textwidth]{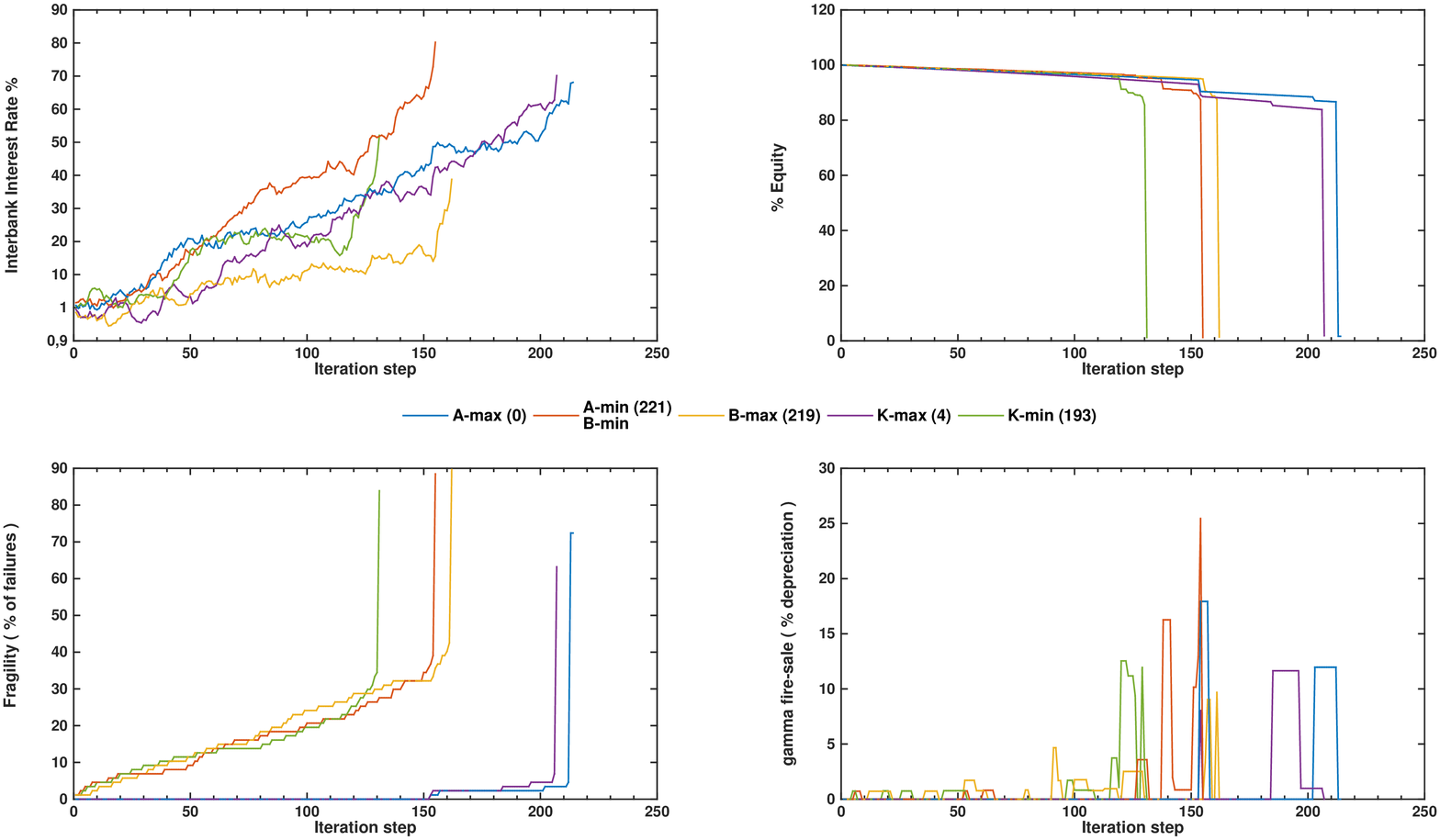}
\caption{Dynamics of a single realization of the ABM built on balance sheet data of year 2004.}
\label{fig:2004}
\end{figure}

\begin{figure}[p!]
\centering
\includegraphics[width=\textwidth]{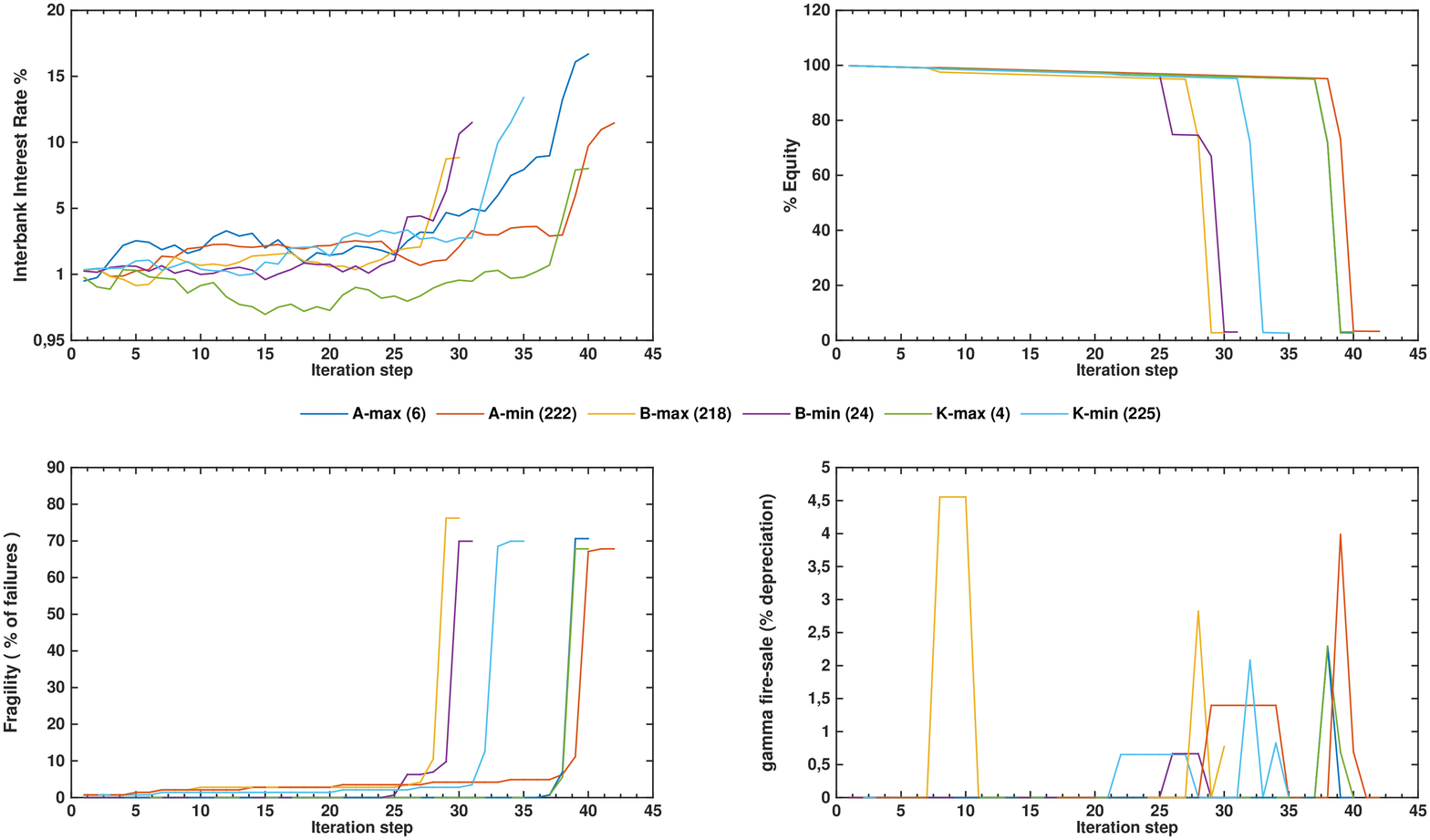}
\caption{Dynamics of a single realization of the ABM built on balance sheet data of year 2008.}
\label{fig:2008}
\end{figure}

\begin{figure}[p!]
\centering
\includegraphics[width=\textwidth]{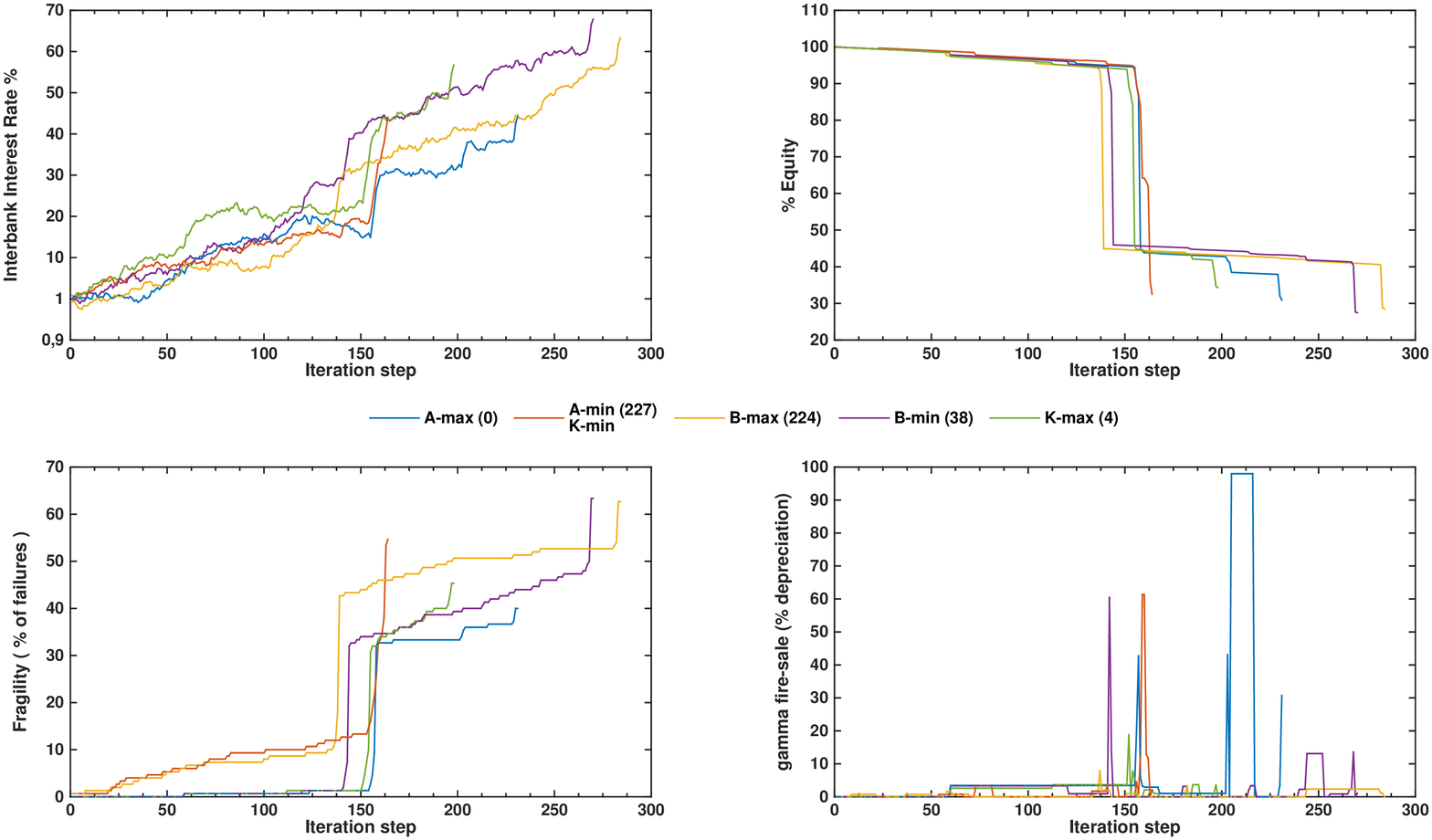}
\caption{Dynamics of a single realization of the ABM built on balance sheet data of year 2013.}
\label{fig:2013}
\end{figure}

We now present results of our ABM simulations (Table \ref{tab.params} reports the list of parameter values we use). 
Firstly, we look at the dynamics of a single realization of the system. Figures \ref{fig:2004}, \ref{fig:2008} and \ref{fig:2013} report results of the model run on balance sheet data 
of some representative years: 2004, 2008 and 2013, respectively. These are the extremal years of the dataset at our disposal plus the year of the global financial crisis\footnote{Results 
for other years are reported in the Supplementary Materials, Figures from \ref{fig:2005} to \ref{fig:2012}.}. 
These figures show various important quantities characterizing the market at different iteration steps $t$: interest rate (upper left panels), percentage of total residual equity (upper right panels), 
percentage of defaulted banks (lower left panels) and the depricing factor $\gamma$ (lower right panels). 
The different trajectories refer to different model configurations, in which we systematically hit a given bank $s$ with the exogenous shock $\Phi$. 
Thus, the line $A$-max ($A$-min) refer to $s$ being the biggest (smallest) bank in terms of total assets; the line $B$-max ($B$-min) to $s$ being the bank with the highest (lowest) leverage; 
the line $K$-max ($K$-min) to $s$ being the bank with most (less) bilateral contracts in the interbank market. 

Looking at the figures, the first striking observation is that the ABM dynamics converge to the market freeze condition much faster in 2008 than in the other years. 
Indeed, the maximum interest rate which can be sustained by the market is also much lower in 2008. Actually, final values of $r$ reached both in years 2004 and 2013 appear to be unreasonably high, 
meaning that the interbank market is rather stable to the proposed dynamics. The total residual equity in the market and the number of defaults have, as expected, a symmetrical trend, 
and the sudden drop of residual equity usually marks the transition to the market freeze state where the assets fire sale depreciation $\gamma$ is maximal. 
Note however a significant difference between the ``stable'' years 2004 and 2013. In the first case, the equity drop ends with the total depletion of the market, just as in 2008. 
In 2013 instead the system can absorb the first systemic crash, falling into a state with non-zero residual equity. Market freeze is not triggered immediately, 
and even when it occurs it does not zero the value of the market. This points to the effectiveness of the new regulatory requirements on banks balance sheets in place after the crisis. 
Concerning the difference of system dynamics between the various shock configurations, we see in general that the convergence of the system to the market freeze is faster 
when the systematically shocked bank is ``small'' (\ie, owning a few total assets and a few contracts, and typically having high leverage). ``Big'' and less leveraged banks are indeed 
more robust to an extensive exogenous shock, but they eventually fail causing the same kind of market transition. The difference in behavior is less evident in 2013, 
suggesting that balance sheets became more homogeneous because of the new regulation. 

\begin{figure}[h!]
\centering
\includegraphics[width=0.75\textwidth]{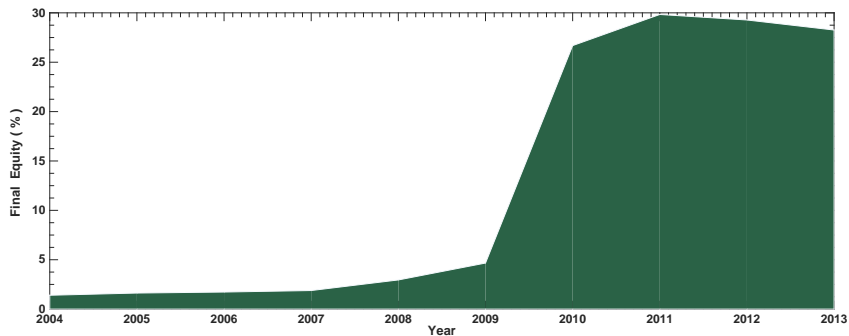}
\caption{Final relative equity in the system after the market freeze, $\sum_iE_i(t_c+1)/\sum_iE_i(0)$, averaged over 1000 ABM run on balance sheet data for different years.}
\label{fig:finalequity}
\end{figure}

\begin{figure}[h!]
\centering
\includegraphics[width=0.75\textwidth]{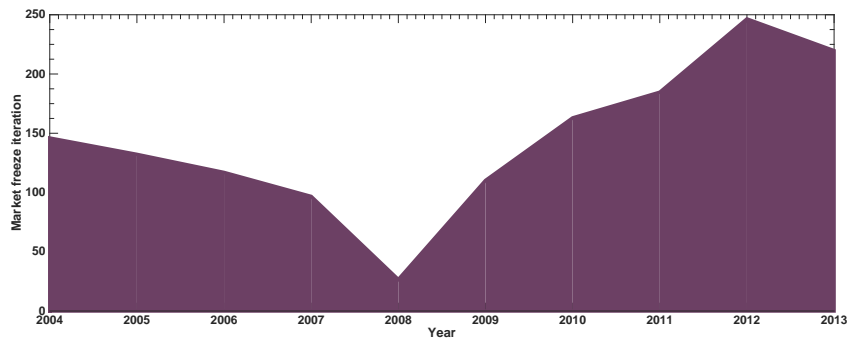}
\caption{Final iteration $t_c$ of the dynamics (market freeze condition), averaged over 1000 ABM run on balance sheet data for different years.}
\label{fig:duration}
\end{figure}

\begin{figure}[h!]
\centering
\includegraphics[width=0.75\textwidth]{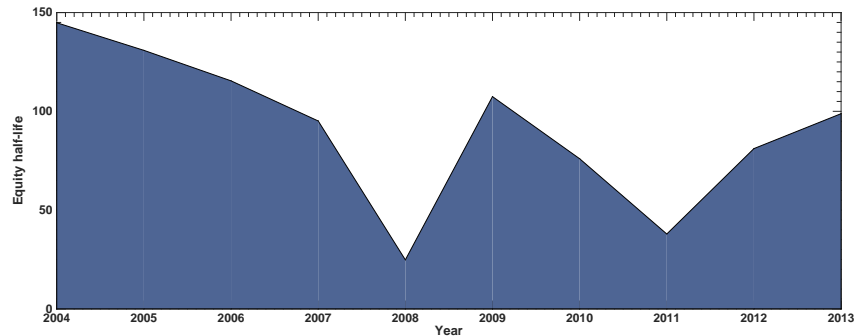}
\caption{Half-life $t_{1/2}$ of the total equity in the market, averaged over 1000 ABM run on balance sheet data for different years.}
\label{fig:timetohalf}
\end{figure}

We further discuss more robust results, which are obtained as averages over 1000 realizations of the ABM dynamics and for distributed exogenous shocks (the bank to be shocked is randomly drawn at each iteration). 
Figure \ref{fig:finalequity} supports our findings outlined above: up to years 2008-2009, the final equity in the system $\sum_iE_i(t_c+1)$ is basically zero, 
whereas, after 2010 we observe a higher resilience of the system, with a residual equity around 30\% even after the freezing of the interbank market. 
Figure \ref{fig:duration} shows instead the length of the ABM dynamics, namely the number of iterations $t_c$ for the system to converge to the market freeze. 
This indicator basically quantifies the maximum delay allowed for a regulatory intervention aimed at taming the crisis spiral. 
Notably, the minimum value of $t_c$ is six times smaller in 2008 than in 2012. We see that the system monotonically loses its resilience before the global crisis, and increases it afterwards. 
Yet, according to our previous analysis, the system converges to its final state in different ways for early and late years of our dataset. 
In particular, the first drop of total equity can well represent the outbreak of the crisis event. We thus introduce the {\em half-life} of the system $t_{1/2}$ as the iteration step at which the total equity 
in the system is halved, \ie, $\sum_iE(t_{1/2})=\tfrac{1}{2}\sum_iE(0)$. In the whole range of years considered, this iteration corresponds to the earliest substantial equity drop. 
As Figure \ref{fig:timetohalf} shows, $t_{1/2}$ behaves differently from $t_c$: there is an additional minimum in 2011 (the year of the European sovereign bond crisis), 
and the more resilient markets are now those far before the global financial crisis. Overall, according to our results, year 2008 marked a transition between a regime 
in which a crisis was hard to be triggered but would lead to a total market crash, and a regime in which a crisis is easier to occur but part of the system is likely to survive it.

\section{Discussion} 

In this work we have designed an agent based model to mimic the dynamics of the interbank lending market during financial crises. 
The model relies on banks balance sheet as the only data source, and is build on simple assumptions on banks strategic behavior during periods of financial distress. 
We find that as we get close to the global financial crisis of 2008, the system becomes less stable in terms of time to collapse. This feature persists after the crisis, 
and another peak of instability is observed in 2011. However, the consequence of a crisis are much more severe---in term of overall losses---before 2009, 
as afterwards the new regulation made banks balance positions more solid. 

Here we focused on the dynamics interbank market because of its crucial role of liquidity provider to the financial system \citep{Allen2014}, and to the economy in general \citep{Gabbi2015}. 
As this system results from the usually uncollateralized (OTC) bilateral contracts between banks, it is rather sensitive to market movements \citep{Smaga2016}, 
and it can dry up under exceptional circumstances \citep{Brunnermeier2009a} becoming a main vehicle for distress spreading in the financial system. 
The dynamics of the interbank market are driven by leverage targeting and liquidity hoarding behaviors of banks. These selfish strategies do consolidate the individual bank position, 
but spread financial distress through spin-off effects like interest rate increase and fire sale spillovers, which in turn induce other banks to adopt similar pro-cyclic behavior. 
Exceptional monetary policies by central banks are usually implemented to sustain the interbank market during periods of deteriorated creditworthiness and distributed distress. 
However, in order to assess the stability of the system, in our model we did not include the possibility of a regulatory intervention nor of bail-outs. 
Indeed, by measuring the speed at which the crisis breaks out, we provide a temporal window for implemented anti-cyclical policies to be effective in mitigating the crisis.

\section*{Acknowledgements}
This work was supported by the EU projects GROWTHCOM (FP7-ICT, grant n. 611272), MULTIPLEX (FP7-ICT, grant n. 317532), SIMPOL (FP7 grant n. 610704), DOLFINS (H2020-EU.1.2.2., grant n. 640772) and CoeGSS (EINFRA, no. 676547). 
The funders had no role in study design, data collection and analysis, decision to publish, or preparation of the manuscript. 

\bibliographystyle{apalike}

\setcounter{table}{0}
\setcounter{figure}{0}
\renewcommand{\thetable}{S\arabic{table}}
\renewcommand{\thefigure}{S\arabic{figure}}

\section*{Supplementary Material}

\begin{figure}[h!]
\centering
\includegraphics[width=\textwidth]{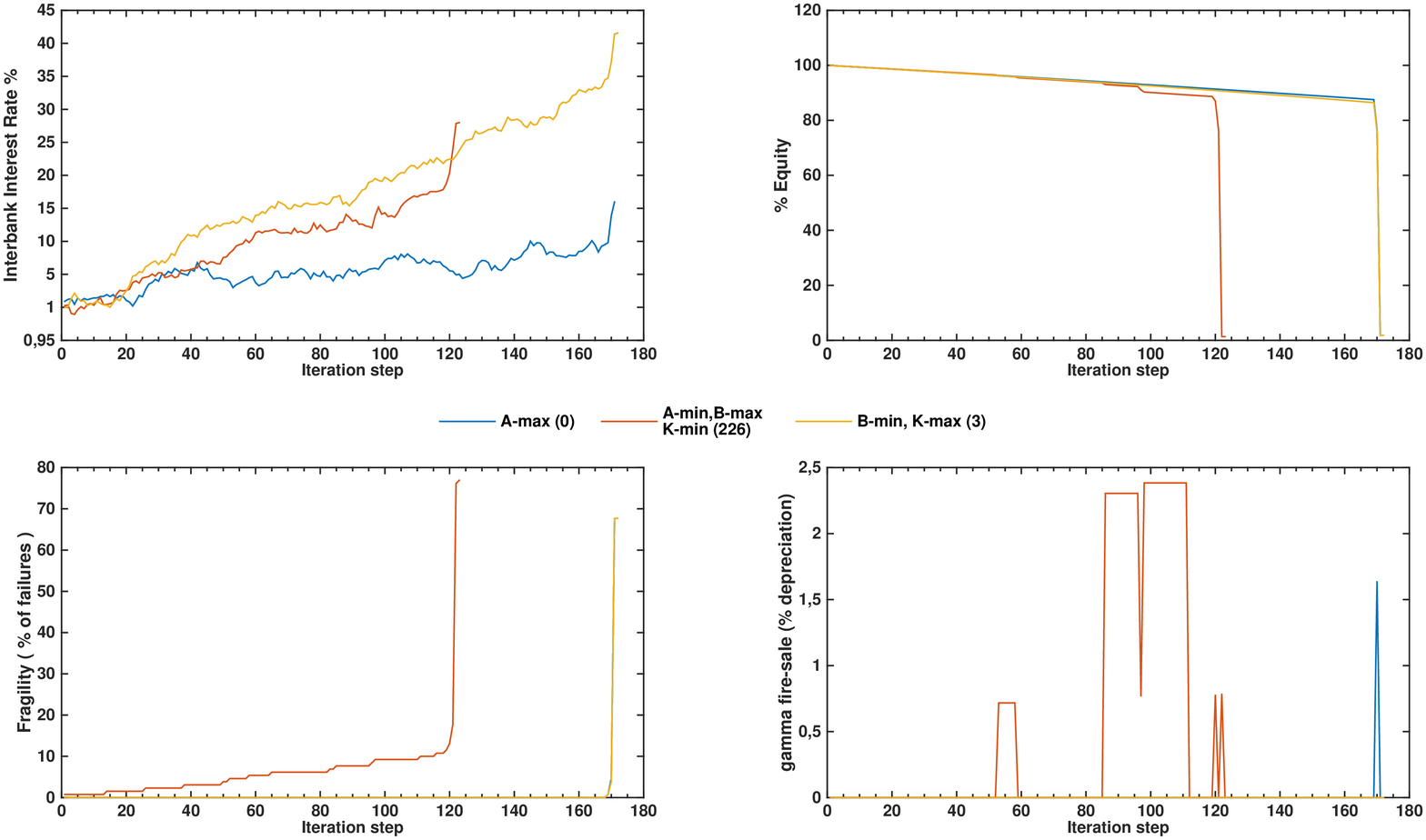}
\caption{Dynamics of a single realization of the ABM built on balance sheet data of year 2005.}
\label{fig:2005}
\end{figure}
\begin{figure}[h!]
\centering
\includegraphics[width=\textwidth]{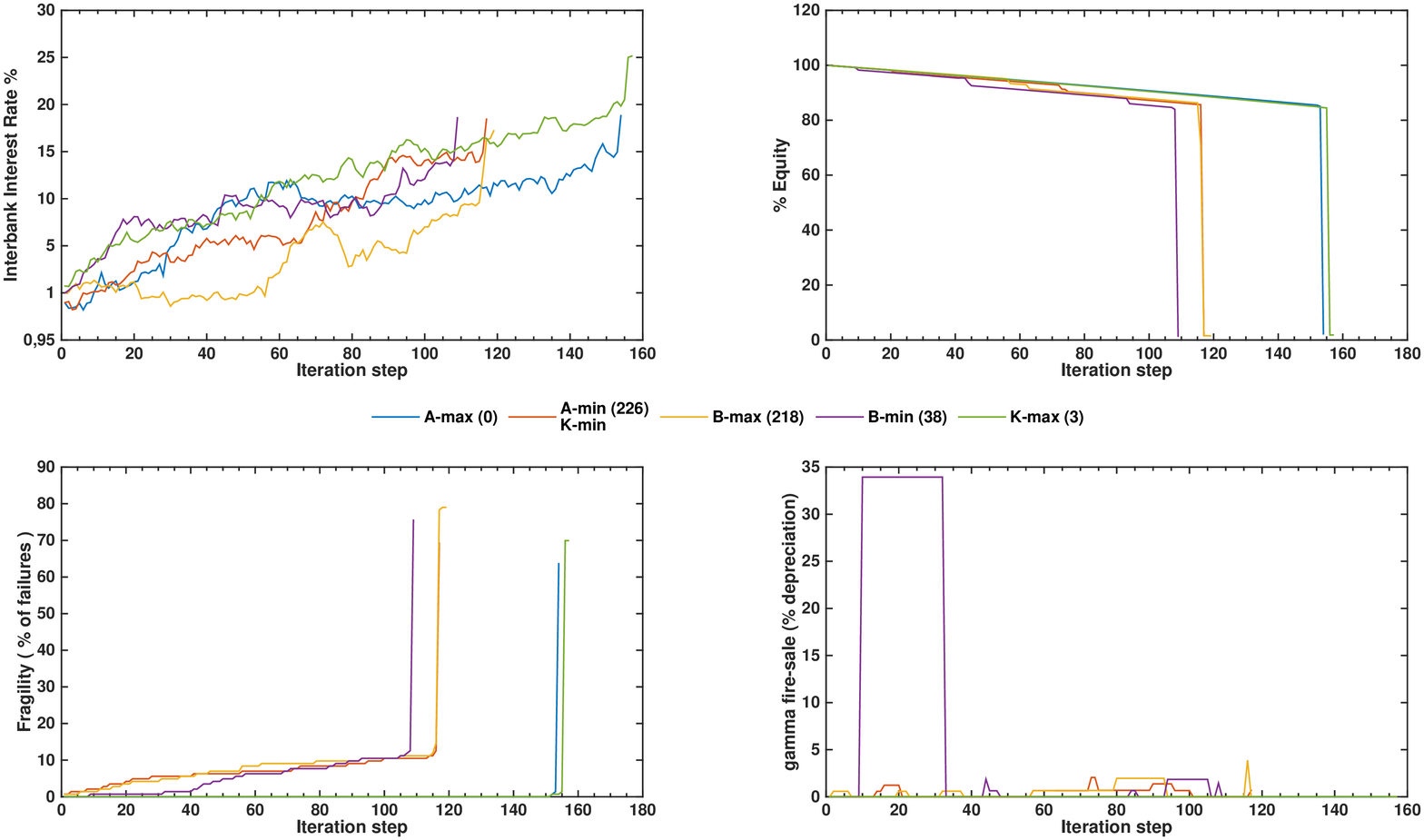}
\caption{Dynamics of a single realization of the ABM built on balance sheet data of year 2006.}
\label{fig:2006}
\end{figure}
\begin{figure}[h!]
\centering
\includegraphics[width=\textwidth]{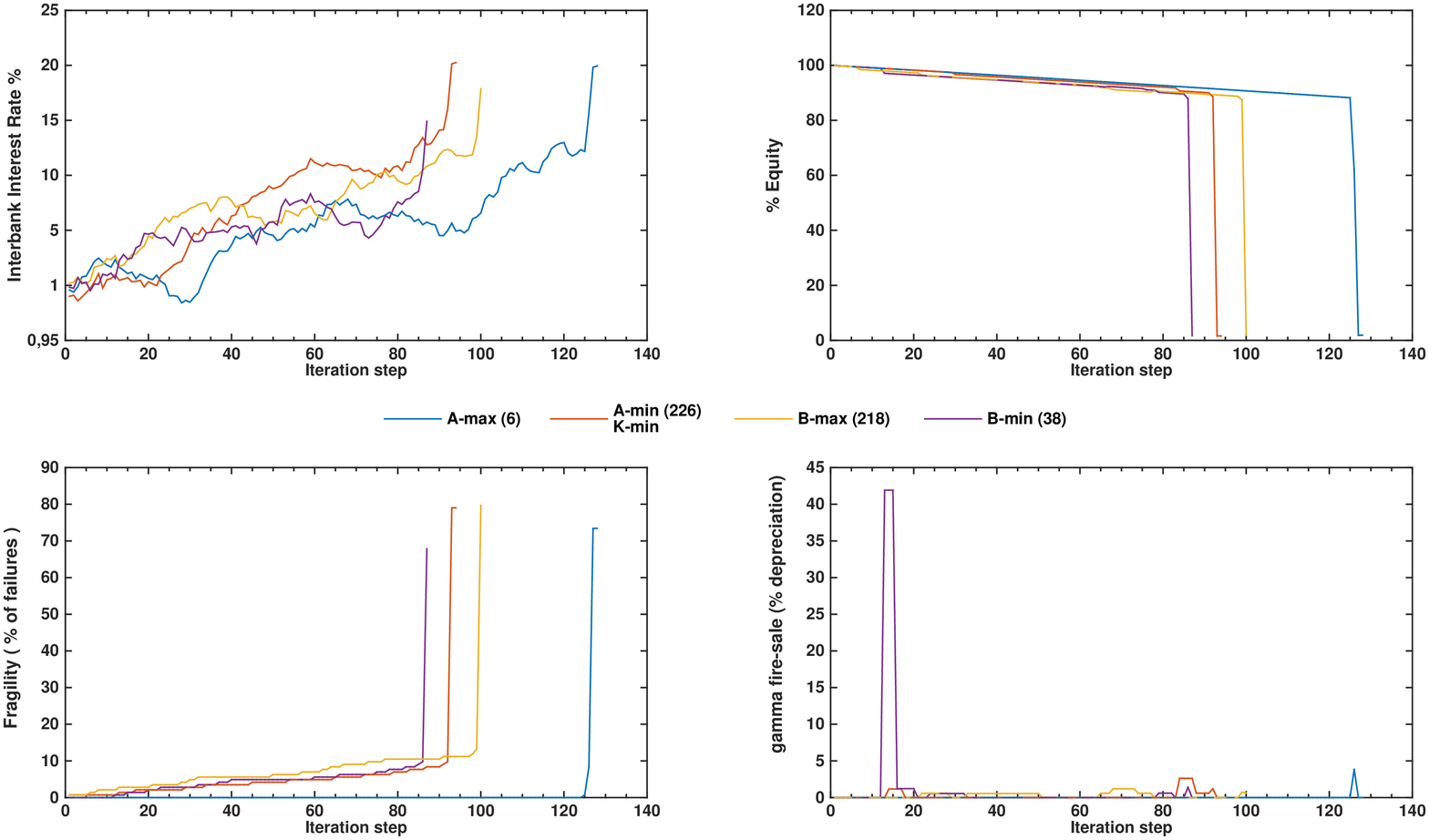}
\caption{Dynamics of a single realization of the ABM built on balance sheet data of year 2007.}
\label{fig:2007}
\end{figure}
\begin{figure}[h!]
\centering
\includegraphics[width=\textwidth]{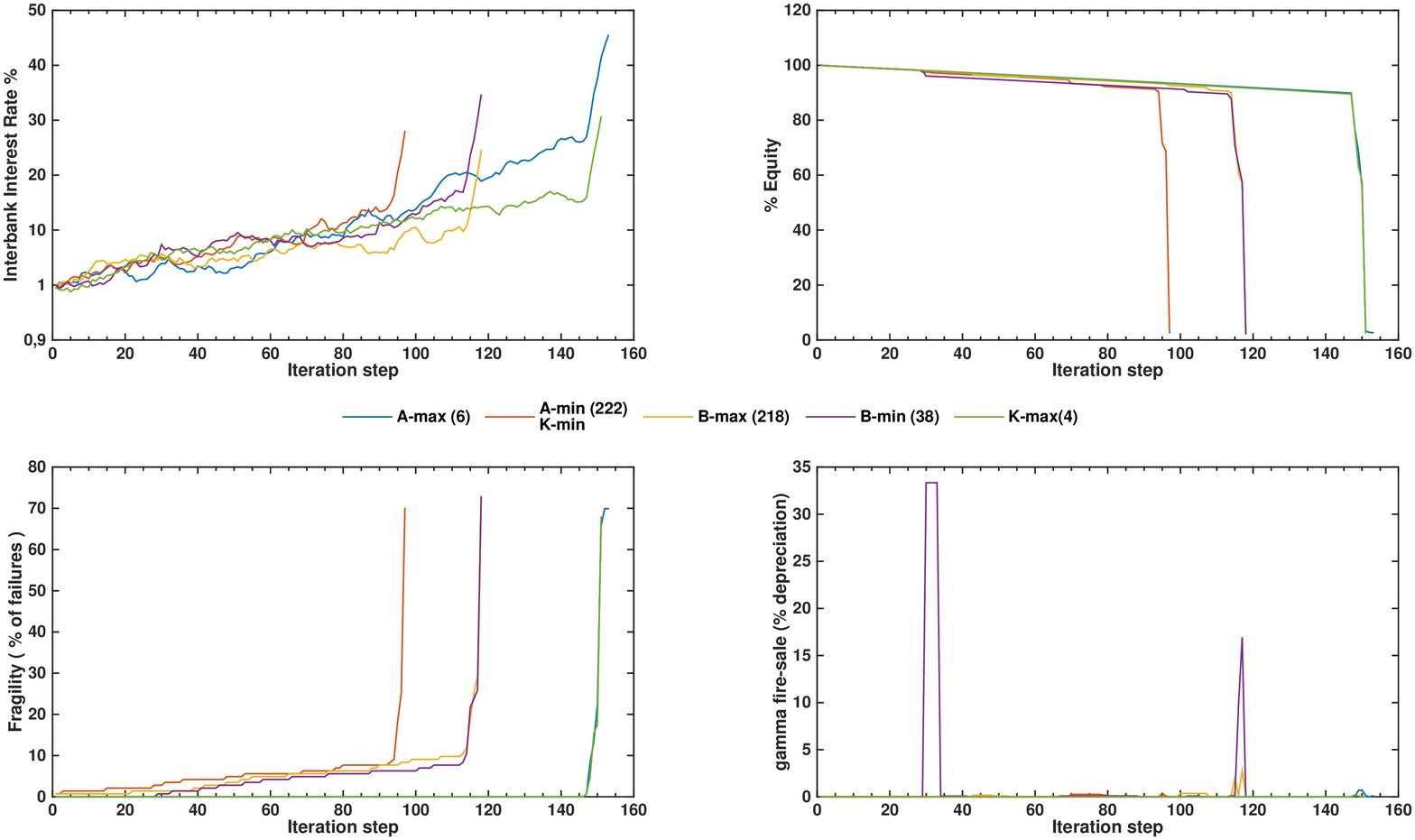}
\caption{Dynamics of a single realization of the ABM built on balance sheet data of year 2009.}
\label{fig:2009}
\end{figure}
\begin{figure}[h!]
\centering
\includegraphics[width=\textwidth]{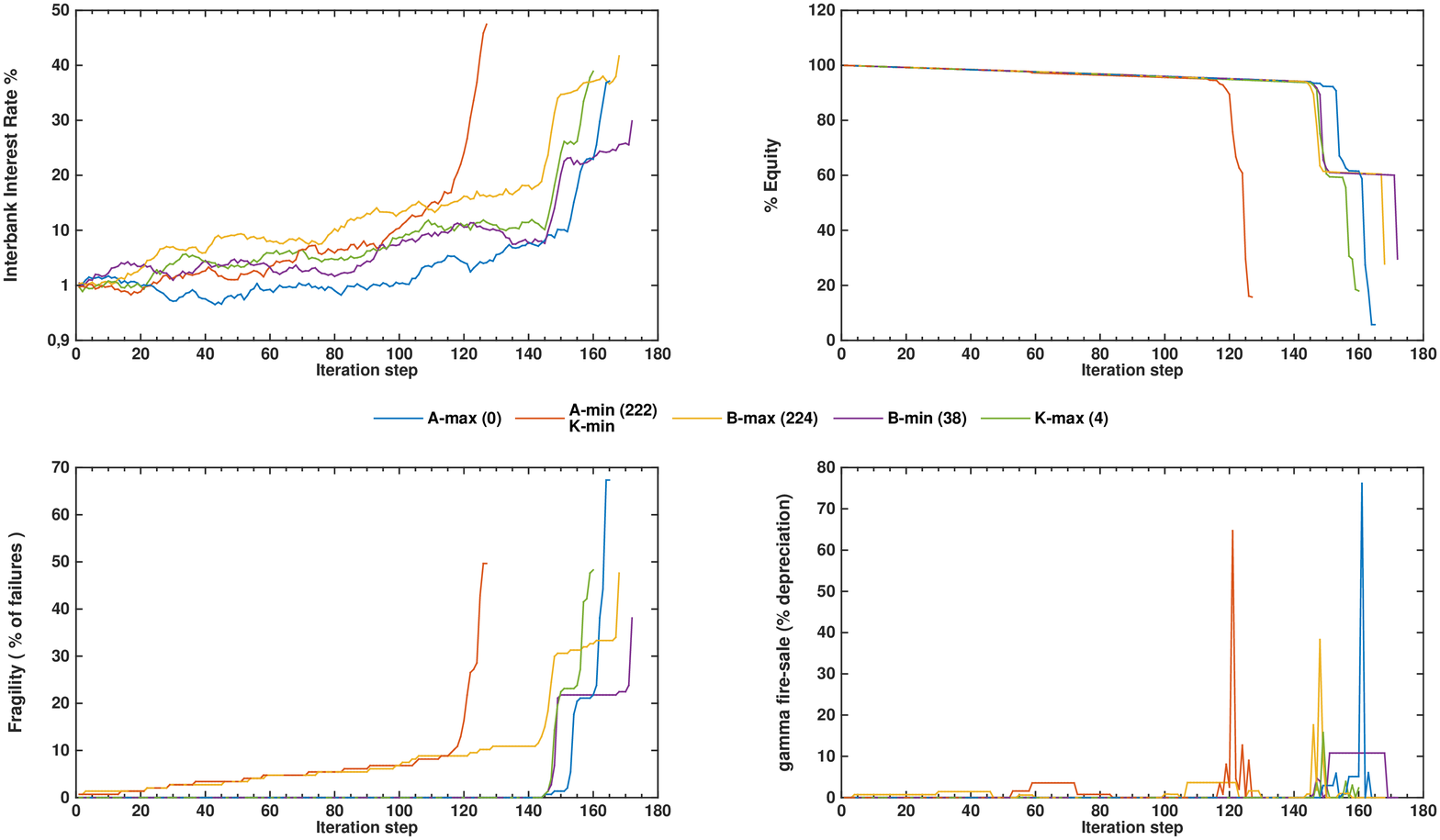}
\caption{Dynamics of a single realization of the ABM built on balance sheet data of year 2010.}
\label{fig:2010}
\end{figure}
\begin{figure}[h!]
\centering
\includegraphics[width=\textwidth]{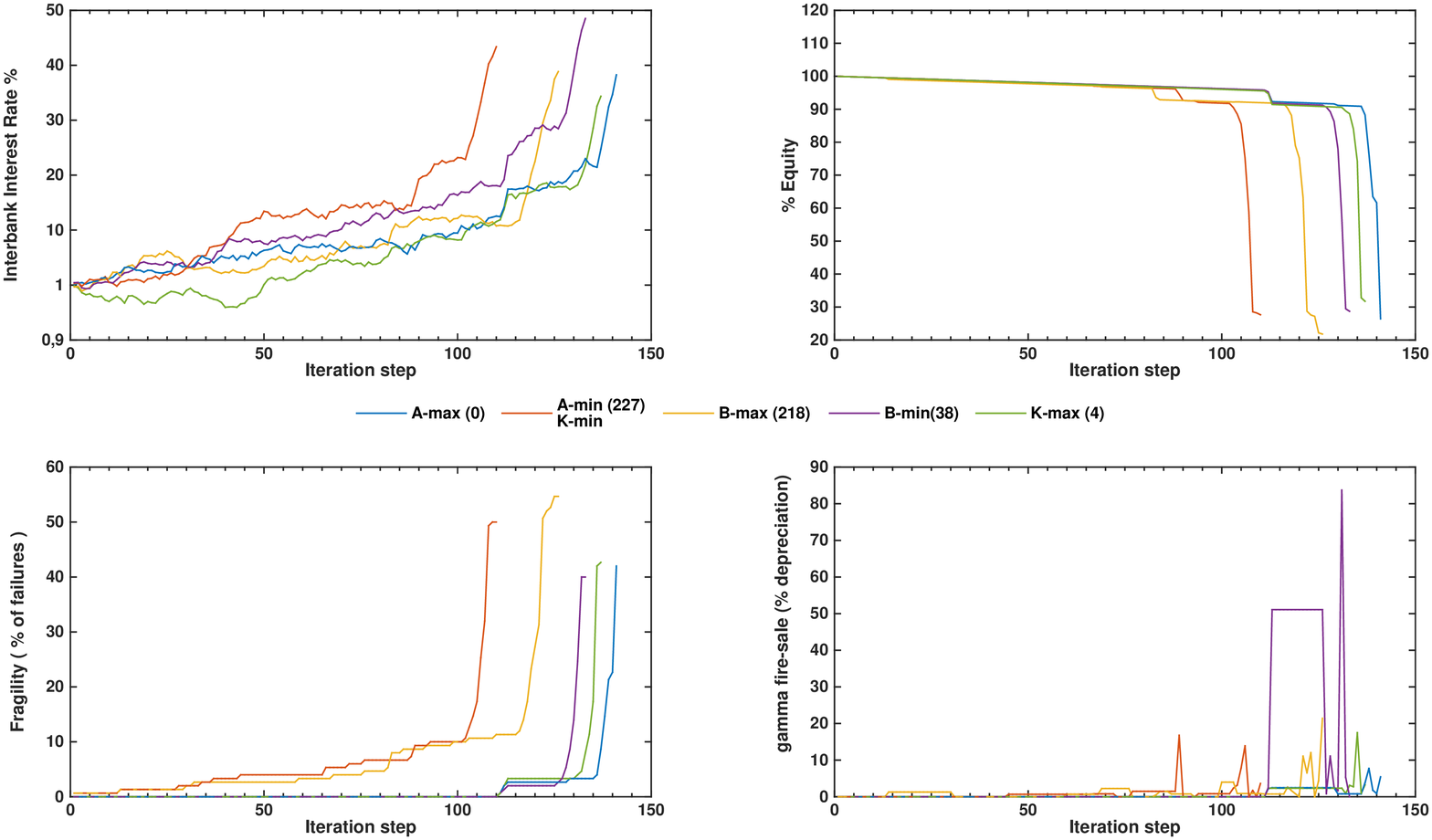}
\caption{Dynamics of a single realization of the ABM built on balance sheet data of year 2011.}
\label{fig:2011}
\end{figure}
\begin{figure}[h!]
\centering
\includegraphics[width=\textwidth]{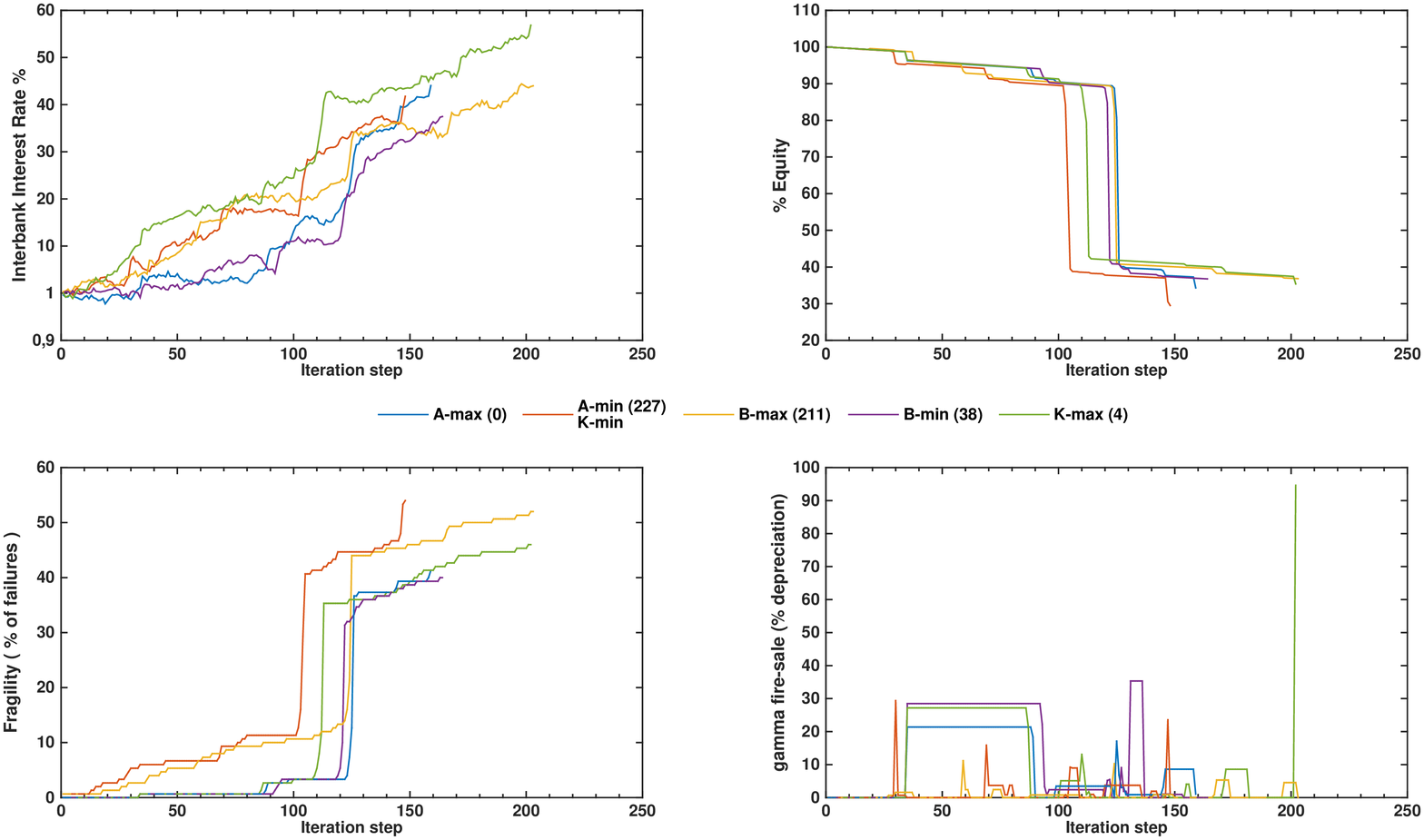}
\caption{Dynamics of a single realization of the ABM built on balance sheet data of year 2012.}
\label{fig:2012}
\end{figure}

\end{document}